\documentclass[%
 aip,
 amsmath,amssymb,
 reprint,%
]{revtex4-1}

\usepackage{graphicx}% Include figure files
\usepackage{dcolumn}% Align table columns on decimal point
\usepackage{bm}% bold math
\usepackage[utf8]{inputenc}
\usepackage[T1]{fontenc}
\usepackage{mathptmx}
\usepackage{etoolbox}

\usepackage[colorlinks=true, allcolors=black]{hyperref}
\usepackage{siunitx}
\usepackage{amsmath}
\usepackage{amssymb}
\usepackage{flushend} % Needed for widetext
\usepackage{xcolor}
\usepackage{nicefrac}
\usepackage{mathtools}

\DeclarePairedDelimiter\ket{\lvert}{\rangle}
\DeclarePairedDelimiterX\braket[2]{\langle}{\rangle}{#1 \delimsize\vert #2}

\makeatletter
\def\@email#1#2{%
 \endgroup
 \patchcmd{\titleblock@produce}
  {\frontmatter@RRAPformat}
  {\frontmatter@RRAPformat{\produce@RRAP{*#1\href{mailto:#2}{#2}}}\frontmatter@RRAPformat}
  {}{}
}%
\makeatother
\begin{document}

\preprint{AIP/123-QED}

% %-------------------------------------------
% % Paper Head
% %-------------------------------------------

\title[]{Low-density InGaAs/AlGaAs Quantum Dots in Droplet-Etched Nanoholes}
\author{Saimon F. Covre Da Silva}
\email{saimon@unicamp.br}
\affiliation{Institute of Semiconductor and Solid State Physics, Johannes Kepler University Linz, Altenberger Straße 69, 4040 Linz, Austria}
\affiliation{Instituto de Física Gleb Wataghin, Universidade Estadual de Campinas (UNICAMP), 13083-859 Campinas, Brazil}
\author{Ailton J. Garcia Jr}
\author{Maximilian Aigner}
\author{Christian Weidinger}
\author{Tobias M. Krieger}
\author{Gabriel Undeutsch}
\affiliation{Institute of Semiconductor and Solid State Physics, Johannes Kepler University Linz, Altenberger Straße 69, 4040 Linz, Austria}
\author{Christoph Deneke}
\affiliation{Instituto de Física Gleb Wataghin, Universidade Estadual de Campinas (UNICAMP), 13083-859 Campinas, Brazil}
\author{Ishrat Bashir}
\affiliation{Department of Electrical Engineering, Indian Institute of Technology Delhi, Hauz Khas, New Delhi, Delhi 110016, India}
\author{Santanu Manna}
\affiliation{Institute of Semiconductor and Solid State Physics, Johannes Kepler University Linz, Altenberger Straße 69, 4040 Linz, Austria}
\affiliation{Department of Electrical Engineering, Indian Institute of Technology Delhi, Hauz Khas, New Delhi, Delhi 110016, India}
\author{Melina Peter}
\author{Ievgen Brytavskyi}
\author{Johannes Aberl}
\author{Armando Rastelli}
\affiliation{Institute of Semiconductor and Solid State Physics, Johannes Kepler University Linz, Altenberger Straße 69, 4040 Linz, Austria}

\date{\today}% It is always \today, today,

% \keywords{Quantum Dots, Local Droplet Etching, Molecular Beam Epitaxy,...}

\begin{abstract}

Over the past two decades, epitaxial semiconductor quantum dots (QDs) have demonstrated very promising properties as sources of single photons and entangled photons on-demand. Among different growth methods, droplet etching epitaxy has allowed the growth of almost strain-free QDs, with low and controllable surface densities, small excitonic fine structure splitting (FSS), and fast radiative decays. Here, we extend the local droplet etching technique to In(Ga)As QDs in AlGaAs, thereby increasing the achievable emission wavelength range beyond that accessible to GaAs/AlGaAs QDs, while benefiting from the aforementioned advantages of this growth method. We observe QD densities of \SI{\sim0.2}{\micro m^{-2}}, FSS values as small as \SI{3}{\micro eV}, and short radiative lifetimes of \SI{\sim300}{ps}, while extending the achievable emission range to \SI{\sim920}{nm} at cryogenic temperatures. We envision these QDs to be particularly suitable for integrated quantum photonics applications.

\end{abstract}

\maketitle

%-------------------------------------------
% Paper Body
%------------------------------------------- 
%

\section{Introduction}
Quantum technologies, especially quantum communication \cite{Ekert1991, Kimble2008, Bennett2014,Wehner2018} and photonic quantum simulation \cite{Zhong2021, Arrazola2021}, require advanced sources of quantum light.
These include sources based on spontaneous parametric down-conversion (SPDC) \cite{Kwiat1999}, trapped atoms \cite{Kuhn2002}, and semiconductor quantum dots (QDs) \cite{Benson2000}. 
QDs are attractive due to their high quantum efficiency and brightness \cite{Ding2025}, as well as near-unity photon indistinguishability \cite{Somaschi2016} and entanglement fidelity\cite{Rota2024}.
They also offer an ``on demand'' operation with sub-Poissonian photon counting statistics \cite{Schweickert2018,Hanschke2018}, while also allowing to fine-tune the emission characteristics across a broad spectral range\cite{Arakawa2020}.
More specifically, InGaAs QDs obtained by the Stranski-Krastanow (SK) growth mode of InAs on GaAs(001) have been used for many pioneering proof-of-principle experiments\cite{Heindl2023} and are now commercially available. These developments have been enabled by the integration of QDs in photonic nanostructures and devices\cite{Lodahl2018}. 
One of the challenges with InGaAs SK QDs is the difficulty of growing them with a surface density low enough to optically address single QDs over a full GaAs wafer\cite{Dubrovsku2004,Kamiya2002,Rastelli2004,Verma2022}. 
In addition, inhomogeneous In alloying leads to large excitonic fine-structure splitting (FSS)\cite{MarFSS} and also a noisy nuclear spin environment\cite{Schimpf2023, Zaporski2023a}, possibly limiting their performance as sources of polarization-entangled photon pairs\cite{Huber2018}. 
Furthermore, the typical excitonic radiative lifetime of \SI{1}{ns}\cite{Vural2020}, as well as the presence of wetting-layer states\cite{Lobl2019}, lead to a significant dephasing, limiting their utility in quantum applications \cite{Lobl2019,Fricker2023}. Although InGaAs QDs with short radiative lifetime (high oscillator strength) can be obtained by InGaAs deposition on GaAs\cite{Reithmaier2004} or by InAs deposition at low growth rates\cite{Kamiya2002} followed by ex-situ rapid thermal processing\cite{Girard2004,Rastelli2004a,Langbein2004}, these approaches are either accompanied by difficulties in controlling the QD density or by an additional high-temperature processing step. 
In contrast, QDs grown by filling local droplet-etched (LDE) nanoholes with GaAs, further simply referred to as GaAs QDs, have overcome some of these limitations. These structures allow for the growth of QDs with higher symmetry and low surface density, improved ensemble homogeneity, and higher oscillator strengths compared to SK QDs\cite{DaSilva2021}.
These superior properties are reflected in a low surface density of approximately $\SI{0.2}{\micro m^{-2}}$, an average FSS of \SI{3}{\micro eV} and below\cite{DaSilva2021} and a short radiative lifetime in the order of \SIrange{200}{250}{ps}\cite{Reindl2019}, making these QDs a promising alternative to SK QDs. When embedded in diode structures, GaAs QDs routinely display emission linewidths close to the transform limit\cite{Zhai2020,Zhai2022,Undeutsch2025}.
However, the longest wavelength achievable by GaAs QDs is inherently limited to the emission wavelength of free excitons in GaAs (\SI{815}{nm} at typical cryogenic temperatures around \SI{5}{K}).
Longer wavelengths are desirable for QDs embedded in nanophotonic structures, where the impact of fabrication imperfections, as well as scattering and absorption losses in Al(Ga)As scale with the wavelength\cite{Michael2007}. 
In addition, experimental setups and technologies designed for conventional InGaAs SK QDs would benefit from QDs emitting in similar wavelength ranges.

Very recently, the benefits of LDE to obtain nanostructures with low surface density have been combined with SK growth to obtain SK QDs with low density over full wafers\cite{Yu2019,Kersting2025}. 
In this work, we take a different approach. Based on an established method for etching nanoholes in AlGaAs using molecular beam epitaxy (MBE), we deposit a layer of In$_x$Ga$_{1-x}$As with a nominal In content of $x=$0.1-0.4 to fill the nanoholes. Surface characterization by atomic force microscopy (AFM) shows a low QD density (\SI{\sim0.22}{\micro m^{-2}}), suitable for single-photon spectroscopy. Optical \text{$\mu$}-photoluminescence (\text{$\mu$}-PL) measurements show very low FSS values (as small as \SI{3}{\micro eV}), radiative lifetimes (around \SI{300}{ps}), and emission linewidths (\SI{13}{\micro eV}), that are comparable to the values reported for GaAs QDs in non-diode structures. Hence, at least for experiments and applications for which strain and alloy disorder are not important, we can maintain the beneficial properties of GaAs QDs and extend the emission wavelength range from below the GaAs bandgap to values usually reachable by InGaAs QDs either treated with partial capping and annealing\cite{Wang2006} or by rapid thermal
treatment\cite{Babiński2001,Girard2004}.

\begin{figure*}[tb]
    \centering
    \includegraphics[width=0.8\textwidth]{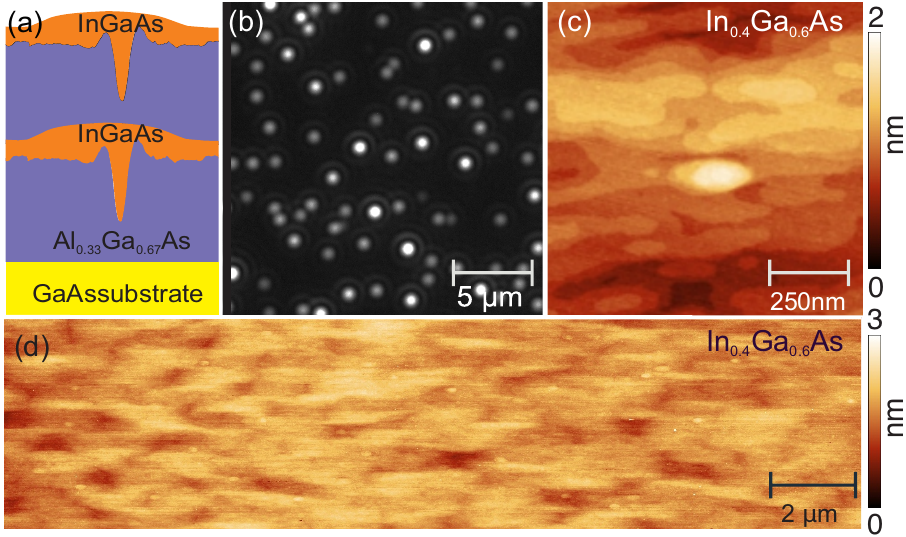}
    \caption{%
    (a) Schematic illustration of the sample structure using AFM linescans (the vertical scale of the nanoholes and mounds is exaggerated).
    (b) Cryogenic PL microscopy image, obtained by illumination of the sample with $x=0.3$ with incoherent blue light, showing the PL of single QDs with a density of \SI{\sim0.22}{\micro m^{-2}}.
    (c) AFM image illustrating the morphology of a single AlGaAs nanohole filled by depositing \SI{1}{nm} of In$_{0.4}$Ga$_{0.6}$As followed by \SI{30}{s} annealing.
    (d) 20 $\times$ 5 $\SI{}{\micro m^2}$ overview AFM image of the same sample as in (c), showing no evidence of SK QD formation.}
    \label{AFM}
\end{figure*}

\section{Sample Growth}
All studied samples were grown on Si-doped GaAs (001) substrates using a solid-source molecular beam epitaxy system (MBE-Komponenten GmbH) equipped with an As-cracker source.
Figure \ref{AFM}(a) shows a schematic illustration of the sample structure. 
Initially, a buffer layer is grown on top of the GaAs substrate at \SI{590}{^\circ C}, followed by a \SI{200}{nm} thick Al$_{0.33}$Ga$_{0.67}$As layer that serves as the bottom barrier for the QDs.
In the following step, we employ LDE to create \SI{\sim 8}{nm} deep nanoholes with Al droplets. A more detailed description of the LDE process and GaAs QD fabrication can be found in Ref.~\cite{DaSilva2021} and in the supplementary information.
The substrate is then cooled to \SI{495}{^\circ C} to limit In desorption \cite{Henini1996} and
\SI{1}{nm} of nominal In$_x$Ga$_{1-x}$As is deposited, followed by a \SI{30}{s} annealing step, a 0.5~monolayers (ML) GaAs cap and another \SI{200}{nm} layer of Al$_{0.33}$Ga$_{0.67}$As, serving as the top barrier.
Subsequently, the etching and filling process is repeated on the surface for AFM measurements. Note that the surface QDs are optically inactive due to the presence of a large density of non-radiative centers. 
Figure \ref{AFM}(b) shows a PL image collected at 10 K of the sample with $x=0.3$. Individual QDs are visible as well-separated bright spots. From these and similar images, we estimate that the QD density is approximately \SI{0.22(3)}{\micro m^{-2}}, enabling single QD spectroscopy. Similar densities, which are solely determined by the initial Al droplet etched nanoholes, are observed also in the other samples discussed in this work. 
Achieving this distribution is straightforward using the LDE method compared to the SK method.
Figure \ref{AFM}(c) shows a high-resolution 1 $\times$ 1 $\SI{}{\micro m^2}$ AFM scan of the sample with $x=0.4$, clearly revealing a filled QD structure after \SI{1}{nm} of InGaAs deposition.
In the surrounding of the formed mound, the surface is characterized by 2D terraces typical of the growth of an InGaAs layer through MBE \cite{Zhou2013}.
The material diffuses across the surface and fills the nanoholes completely.  The mound is elongated along the [1–10] direction, with an approximate length of \SI{300}{nm}, a width of \SI{120}{nm} and a height \SI{1.5}{nm}.
The mound formation and evolution arise from the anisotropic diffusion of material on top of the AlGaAs surface \cite{Rastelli2004}, which has the tendency to fill the etched holes according to their morphology \cite{DaSilva2017}. 
Figure \ref{AFM}(d) represents a 20 $\times$ 5 $\SI{}{\micro m^2}$ AFM overview image of the same sample (with $x=0.4$), showing no evidence of SK QD formation over a large area. The bright spots correspond to the mounds formed after nanohole filling, as shown in Figure \ref{AFM}(c). The absence of SK island formation in the current and related samples is attributed to the low amount of deposited InGaAs (\SI{1}{nm}), which is below the critical thickness required for 3D island formation for the used In concentrations (approximately \SI{2}{nm} for In$_{0.4}$Ga$_{0.6}$As / Al$_{0.33}$Ga$_{0.67}$As\cite{Wang2019}).

\section{Optical Measurements}
\begin{figure}[tb]
    \centering
    \includegraphics[width=\columnwidth]{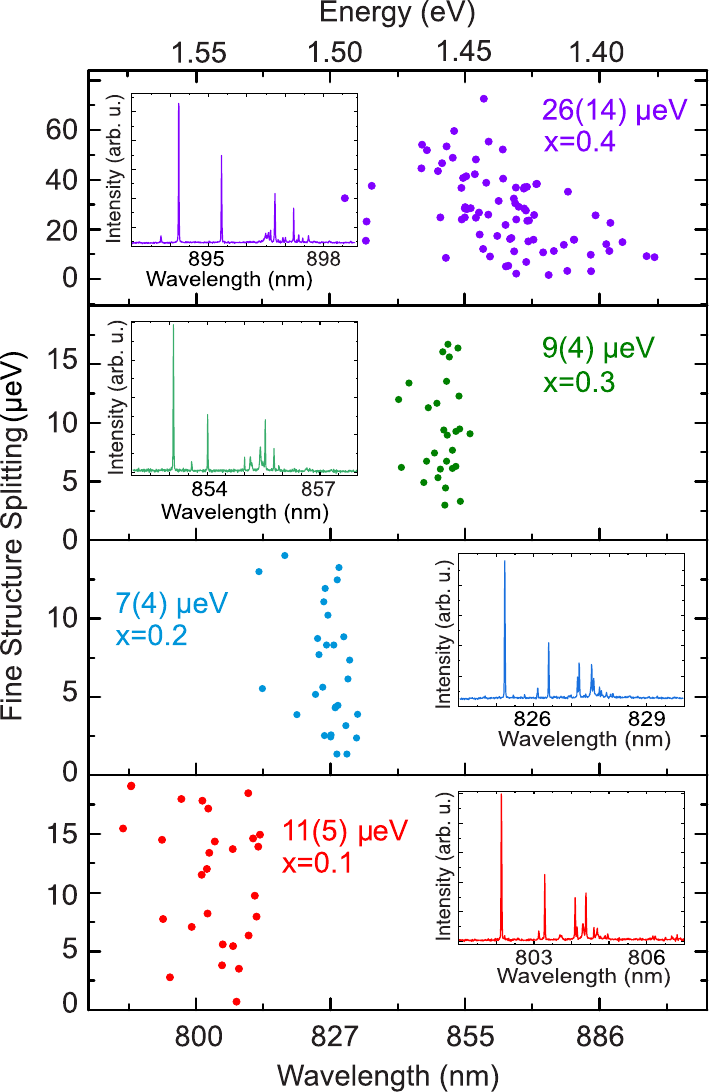}
    \caption{%
    Scatter plots of excitonic FSS and emission wavelength of InGaAs QDs obtained by filling AlGaAs nanoholes by deposition of In$_{x}$Ga$_{1-x}$As with different values of $x$. From the bottom to the top, panels show results for $x=$~0.1, 0.2, 0.3, 0.4. The inset in each panel shows a representative spectrum of an isolated single QD with the respective nominal In concentration.}
    \label{fssvswl}
\end{figure}

For the basic optical characterization of single QDs, a confocal \text{$\mu$}-PL setup with a 50$\times$ microscope objective with 0.42 NA is used. Measurements are performed with samples at \SI{10}{K} and with above bandgap excitation (\SI{533}{nm} diode laser). A separate confocal \text{$\mu$}-PL setup with a 0.65 NA objective and a tunable, pulsed titanium-sapphire laser is employed for time-correlated single-photon counting measurements, coherence time measurements, and auto-correlation measurements. These results are presented in Figures 3--5. 

Figure \ref{fssvswl} shows the result of \text{$\mu$}-PL measurements on randomly chosen single QDs. Compared to GaAs QDs, the emission wavelength of the neutral exciton in these QDs is red-shifted. This shift increases with the In content in the QD, allowing for emission wavelength control from \SI{780}{nm} up to \SI{920}{nm} by varying the nominal In concentration during growth. 

For each $x$ value, a representative emission spectrum is shown in the respective inset. The spectra qualitatively resemble those of GaAs QDs \cite{DaSilva2021} with the dominant emission stemming from the neutral exciton (X) recombination and additional lines at longer wavelengths.
For the samples with $x=$~0.1, 0.2, and 0.3, 30 QDs were measured, showing a wavelength distribution narrower than \SI{30}{nm}. On the sample with $x=0.4$, a higher spread in wavelength distribution of above \SI{45}{nm} was observed, so 80 QDs were measured for better statistics.
Most of the observed QDs have a X linewidth below the setup resolution limit of about $\SI{0.02}{nm}$.

To gather information on the alloy composition of the obtained QDs, we estimated the ground-state optical transitions of InGaAs-filled nanoholes by approximating these QDs as \SI{8}{nm} thick In$_{x}$Ga$_{1-x}$As quantum wells (QWs) with Al$_{0.33}$Ga$_{0.67}$As barriers and by calculating the confined electronic energy levels within the envelope function approximation using single band and 8-band k·p calculations. The calculated transition energies were found to be significantly red-shifted in comparison to the experimentally measured values, suggesting that the actual In incorporation within the QDs is lower than the nominal values set in the growth recipe. 
As an example, for the sample with nominal In fraction $x=$~0.4, we experimentally find an emission wavelength near \SI{900}{nm}, which corresponds to a QW with an $x$ of only 0.15. This discrepancy likely arises from In surface segregation leading to a lower than nominal incorporation of In into the nanoholes and a net loss of In during the temperature ramp step preceding the growth of the AlGaAs layer on top of the InGaAs layer. Further details on the model and growth dynamics are provided in the supplementary information.

Next, the excitonic FSS is assessed via polarization-resolved measurements.
The distribution of FSS has average values of $\SI[separate-uncertainty = false]{11(5)}{\micro eV}$ for $x=$ 0.1, $\SI[separate-uncertainty = false]{7(4)}{\micro eV}$ for $x=$ 0.2, $\SI[separate-uncertainty = false]{9(4)}{\micro eV}$ for $x=$~0.3, and $\SI[separate-uncertainty = false]{26(14)}{\micro eV}$ for $x=$ 0.4. We note here that within the sample with $x=0.4$ we also find QDs with considerably smaller FSS of less than $\SI{3}{\micro eV}$. In general, this sample exhibits a broader distribution both in wavelength and FSS compared to the other samples.
Although the spread in emission wavelength of InGaAs/AlGaAs QDs is larger than in GaAs QDs, the ordering of excitonic complexes is comparable, which is usually not the case in SK QDs.\cite{Trotta2013}

\begin{figure}[tb]
    \centering
    \includegraphics[width=\columnwidth]{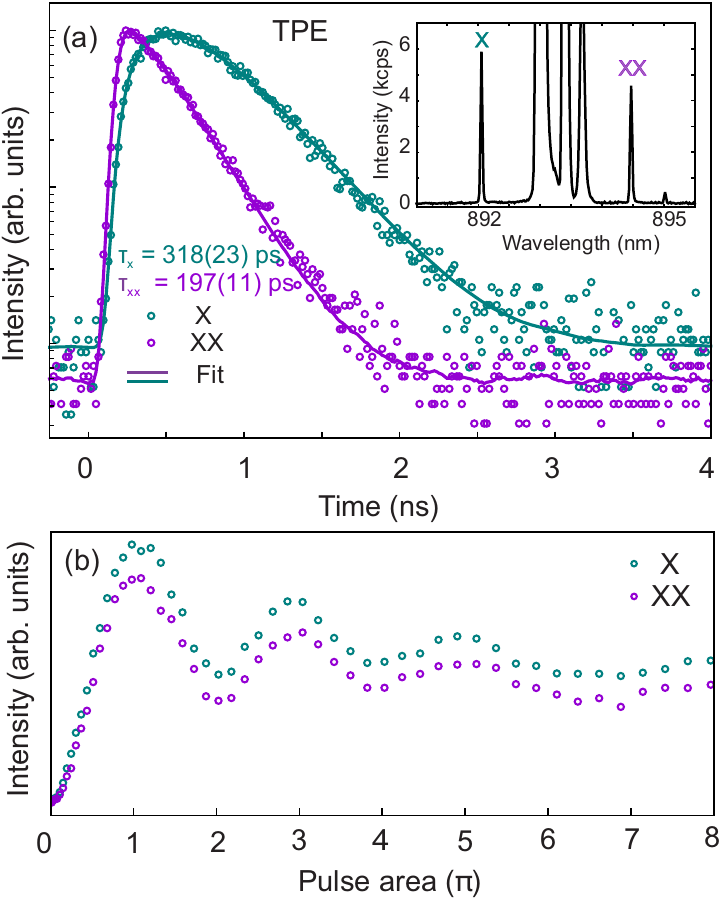}
    \caption{%
    (a) Time-correlated-single-photon-counting measurement of the decay of $\ket{\text{X}}$ (green) and $\ket{\text{XX}}$ (purple) confined in a QD in the sample with $x$ = 0.4 after coherent TPE through a laser with $\approx\SI{893}{nm}$ wavelength, <~\SI{10}{ps} pulse duration, and \SI{80}{MHz} repetition rate. The inset shows the corresponding PL spectrum (residual laser light is seen in between the X and XX lines).
    (b) Excitation-power-dependent behavior of peak intensity versus the excitation pulse area for X (green) and XX (purple) emission for the same QD, showing clear Rabi oscillation.}
    \label{rabi}
\end{figure}

To gain further insight into the optical properties of our InGaAs/AlGaAs QDs, we investigate the sample with $x=0.4$ using different excitation schemes. Pulsed two-photon resonant excitation (TPE) enables coherent population of the biexciton state ($\ket{\text{XX}}$) followed by the emission of XX and X photons via a radiative cascade \cite{Muller2014}.
For charge stabilization, we use an additional continuous wave laser with above-bandgap energy.
The inset in Figure~\ref{rabi}(a) displays a spectrum showing XX and X emission lines, as well as partially filtered laser light in between.
In Figure~\ref{rabi}(b), the power dependence of the intensity of XX and X lines is shown, exhibiting well-defined Rabi oscillations, confirming the coherent control of the $\ket{\text{XX}}$ state. 
Figure~\ref{rabi}(a) shows the results of time-correlated single-photon counting measurements. Lifetimes $\tau_{\text{XX,TPE}} =\SI{197(11)}{ps}$ and $\tau_{\text{X,TPE}} = \SI{318(23)}{ps}$ are obtained by performing a single- and double-exponential fit of the data, respectively, convoluted with the instrumental response function.
The values are approximately three times shorter than those measured in SK-grown InGaAs QDs\cite{Vural2020} emitting in the same wavelength range and are comparable to SK-QDs treated with rapid thermal annealing \cite{Braun2016}. 

\begin{figure}[tb]
    \centering
    \includegraphics[width=\columnwidth]{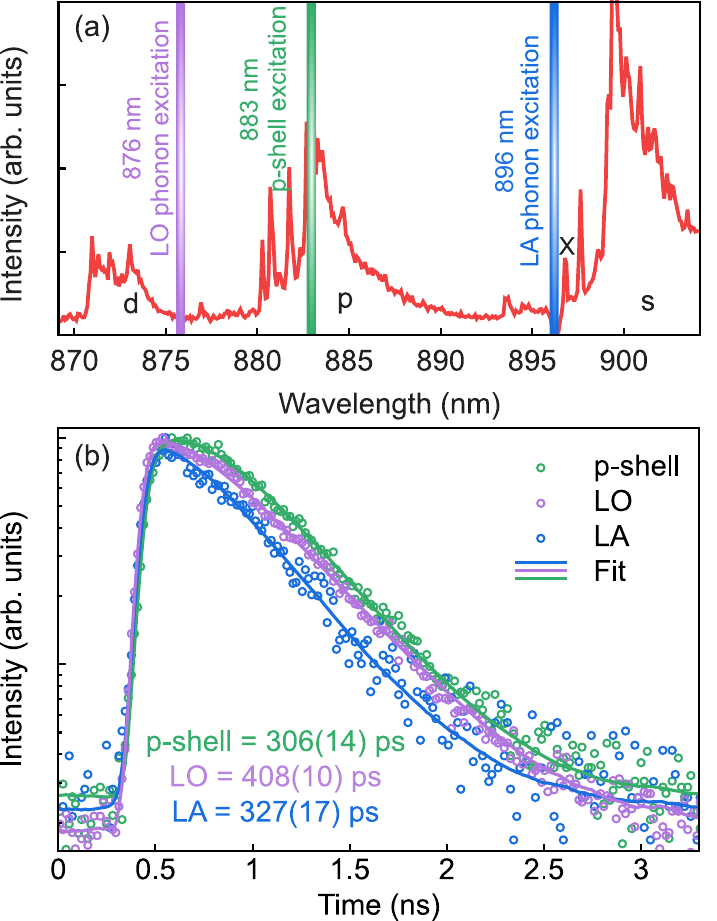}
    \caption{%
    (a) PL spectrum of a QD in the sample with $x=0.4$ under high-power above-bandgap excitation, showing the position of the s-, p-, and d-shell.
    (b) Time-correlated single-photon counting conducted under various excitation schemes (LA-phonon, p-shell and LO-phonon). The measurements were performed on X photons (\SI{897}{nm}).}
    \label{excitation}
\end{figure}

To provide insights into carrier capture and relaxation times via phonon-mediated processes, we extend time-correlated single-photon counting of X photons to incoherent excitation techniques. In fact, it is known that GaAs QDs obtained by the LDE method are characterized by slow interlevel relaxation, which often masks the true radiative decay when excitation is performed using laser energies above-bandgap or resonant to QD excited states\cite{Reindl2019}.
Furthermore, incoherent excitation is easier to implement than resonance fluorescence because of straightforward spectral laser filtering.
To illustrate the QD level structure, essential for optimizing excitation schemes, a spectrum under high-power above-bandgap excitation of a representative QD with $x=0.4$ is shown in Figure \ref{excitation}(a).
From the spectral positions of the different emission bands, an energy separation of approximately \SI{26}{meV} (\SI{16}{nm}) between s-shell and p-shell can be extracted. This value is approximately twice what is observed in common GaAs QDs with emission wavelength around 780~nm\cite{Reindl2019}. Since the AlGaAs nanoholes are fabricated following the same procedure and the amount of material used for nanohole filling is also similar, we expect the sizes and shapes of the QDs studied here to be similar to those of GaAs QDs. We thus attribute the difference in shell spacing to the presence of In, which not only decreases the average energy bandgap of the QD material, but also increases the conduction-band offset and decreases the carrier effective masses, thus increasing the particle confinement energies. 
 
Commonly used incoherent excitation techniques are longitudinal acoustic phonon-assisted (LA), p-shell, and longitudinal optical phonon-assisted (LO) excitation. We employ these by tuning a pulsed laser to the respective wavelengths shown in Figure \ref{excitation}(a).
For each excitation wavelength, we record time traces, shown in Figure \ref{excitation}(b), which are fitted to extract the total lifetime of the excited state. 
For LA excitation, we expect the phonon interaction to be fast compared to the lifetime of the exciton. Data are therefore well fitted through a single exponential fit, resulting in $\tau_{X,LA} = \SI{327(17)}{ps}$.
For ``p-shell'' excitation, carriers first need to relax to the s-shell before the X emission can occur. In general, this relaxation time can not be neglected and is taken into account by performing a double exponential fit. 
We extract a relaxation time into the s-shell of \SI{255(18)}{ps} and an exciton lifetime of $\tau_{X,p} = \SI{306(14)}{ps}$. Ideally, LO excitation also leads to a single exponential decay of the X photon counts. From the fit, we extract a lifetime of $\tau_{X,LO}=\SI{408(10)}{ps}$, which is however larger than the values obtained by the other excitation techniques. We assume that additional states are excited by the laser, which lead to a mixture of different decay paths. The delayed decay we observe compared to the data obtained under LA-assisted excitation is consistent with this interpretation.  
A comparison to GaAs QDs indicates that the transition from p- to s-shell is significantly faster for the InGaAs QDs studied here\cite{Jahn2016,Reindl2019}. We ascribe this observation to the dependence of the relaxation time on the s-p energy separation\cite{Grange2007,Zibik2009}. 

\begin{figure}[tb]
    \centering
    \includegraphics[width=\columnwidth]{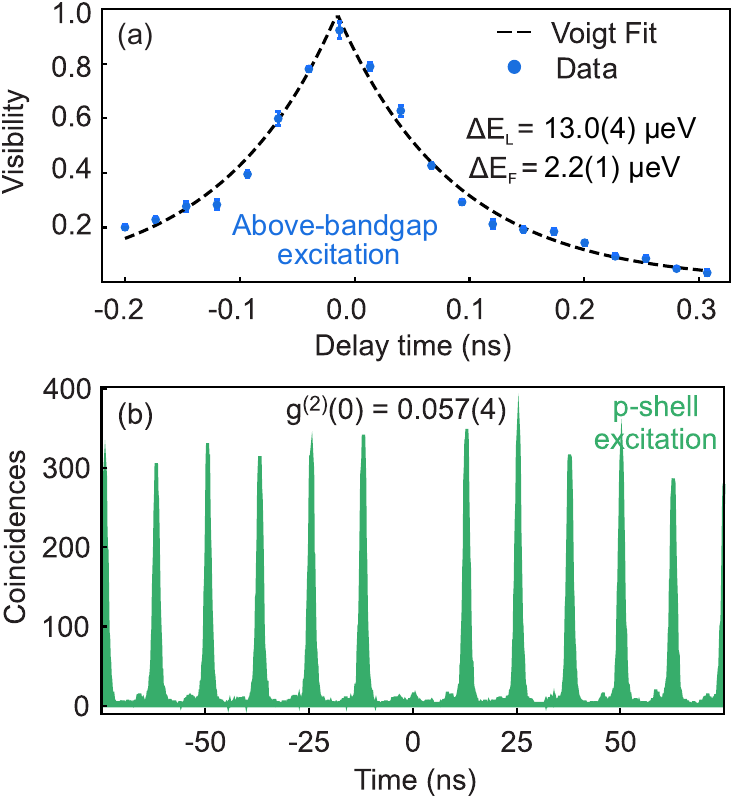}
    \caption{%
    (a) Interference fringe contrast of neutral exciton photons from a QD in the sample with $x=0.4$ as a function of time delay between the two branches of a Michelson interferometer under above-bandgap excitation at 10~K. The Fourier limit $\Delta E_F$ was estimated from the lifetime obtained through p-shell excitation.
    (b) Autocorrelation histogram recorded using the X emission of another QD in the same sample under p-shell excitation at 10~K.}
    \label{decay_michelson}
\end{figure}

To further assess the optical quality of our QDs, we determine the time-averaged coherence time of X photons emitted by a single QD in the sample with $x$ = 0.4 under above-bandgap excitation. To this end, the interference visibility was measured in a Michelson interferometer for different relative time delays among the two interferometer arms. The results, shown in Figure \ref{decay_michelson}(a), are fitted with the Fourier transform of a Voigt profile. We extract a linewidth of $\Delta E_{\text{L}}=\SI{13.0(4)}{\micro eV}$, which, compared to the Fourier limit ($\Delta E_{\text{F}}=\SI{2.2(1)}{\micro eV}$) is broadened by a factor of \SI{5.9(5)}{}. This value is comparable to that typically observed in state-of-the-art GaAs QDs and is likely dominated by charge noise, which can be mitigated by embedding the QDs in a p-i-n diode structure \cite{Zhai2020}. 

Eventually, the single photon emission characteristic was assessed using a Hanbury-Brown-Twiss (HBT) setup.
The resulting histogram, collected under p-shell excitation, is shown in Figure \ref{decay_michelson} (b). Clear antibunching can be observed and a $g^{(2)} (0)$ value of 0.057(4) is obtained. We attribute this finite value to the background of the laser used for charge stabilization, as well as laser reflections on optical elements, that are also visible as mounds between main correlation peaks.

\section{Conclusion and Outlook}

In conclusion, we have shown that it is possible to obtain high-quality InGaAs QDs in LDE-etched AlGaAs nanoholes and adjust their emission wavelength by controlling the nominal In concentration, while retaining some of the favorable properties of established GaAs/AlGaAs QDs.
These InGaAs QDs exhibit the same low surface density as those of GaAs QDs, making them ideal for devices based on single QDs. In particular, we expect the longer emission wavelength compared to GaAs QDs to be advantageous for integrated quantum photonics, as longer wavelengths are associated with lower optical losses and relaxed fabrication tolerances. 
Additionally, it may become possible to interface these QDs with Caesium atom-based quantum memories not only via the $D_1$ transitions at \SI{895}{nm}, but also using the $D_2$ lines at \SI{852}{nm}, that are usually inaccessible to conventional InGaAs/GaAs SK QDs.
The QDs presented here also feature small excitonic FSS, which is useful for the generation of polarization-entangled photon pairs. Although the FSS presented in this work, especially for the sample with $x=0.4$, is higher relative to recent works using GaAs QDs\cite{DaSilva2021}, further optimization of the growth process is still possible. 
The observed excitonic radiative decay times of $\sim$ \SI{300}{ps} are close to those observed for GaAs QDs and faster compared to SK grown QDs emitting in the same wavelength range, thus decreasing the impact of dephasing mechanisms.
Compared to GaAs QDs, a key difference is that the InGaAs QDs presented here have a higher s-p shell separation, up to a factor of two.
This could extend high-fidelity entangled photon emission from \SI{4}{K} to temperatures above \SI{40}{K}\cite{Lehner2023}, allowing the operation of QDs with undegraded entanglement using cost-effective Stirling cryocoolers.
Finally, the optical linewidths we measured for InGaAs QDs are good but have not yet reached the Fourier-transform limit. Strategies to further improve the optical quality include replacing the AlGaAs barrier with GaAs, as this can improve optical quality by reducing impurity-related defects, and/or embedding the QDs in diode structures.

\begin{acknowledgments}
This project has received funding from the European Union’s Horizon 2020 research and innovation program under Grant Agreement No. 871130 (Ascent+) and the EU HE EIC Pathfinder challenges action under grant agreement No. 101115575, from the QuantERA II program that has received funding from the European Union’s Horizon 2020 research and innovation program under Grant Agreement No. 101017733 via the projects QD-E-QKD and MEEDGARD (FFG Grants No. 891366 and 906046) the Austrian Science Fund FWF via the Research Group FG5, I 4320, I 4380, from the Austrian Science Fund FWF 42 through [F7113] (BeyondC), and from the cluster of excellence quantA [10.55776/COE1] as well as the Linz Institute of Technology (LIT), and the LIT Secure and Correct Systems Lab, supported by the State of Upper Austria.
S.F.C. da Silva acknowledges the São Paulo Research Foundation (FAPESP Process Numbers 2024/08527-2 and 2024/21615-8) and the Brazilian National Council for Scientific and Technological Development (CNPq), grant number 300636/2025-3 for financial support. Ch. Deneke thanks FAPESP (Process 2023/01517-9) for the grant supporting his stay at the JKU.
\end{acknowledgments}

\section*{Supporting Information}
Supporting Information is available online free of charge and contains:
Growth of quantum dots (representative growth protocol), modeling experimental data of transition energies

\section*{Data Availability Statement}
The data underlying this study are openly available in Zenodo at https://doi.org/10.5281/zenodo.16760367
\section*{References}
\bibliography{references.bib}% Produces the bibliography via BibTeX.

\end{document}